\documentclass[useAMS,usenatbib]{mn2e}
\usepackage{amssymb}
\usepackage{graphicx}

\title[Magnetar spin-down in extended emission SGRBs]{Can magnetar spin-down power extended emission in some short GRBs?}
\author[B. P. Gompertz, P. T. O'Brien, G. A. Wynn and A. Rowlinson]{B. P. Gompertz$^{1}$\thanks{E-mail: bpg6@le.ac.uk}, P. T. O'Brien$^{1}$, G. A. Wynn$^{1}$ and A. Rowlinson$^{2}$\\
$^{1}$Department of Physics and Astronomy, University of Leicester, Leicester, UK, LE1 7RH\\
$^{2}$Astronomical Institute ``Anton Pannekoek'', University of Amsterdam, Postbus 94249, 1090 GE Amsterdam, The Netherlands}

\begin{document}

\date{Accepted: }

\pagerange{\pageref{firstpage}--\pageref{lastpage}} \pubyear{????}

\maketitle

\label{firstpage}

\begin{abstract}
Extended emission gamma-ray bursts are a subset of the `short' class of burst which exhibit an early time rebrightening of gamma emission in their light curves. This extended emission arises just after the initial emission spike, and can persist for up to hundreds of seconds after trigger. When their light curves are overlaid, our sample of 14 extended emission bursts show a remarkable uniformity in their evolution, strongly suggesting a common central engine powering the emission. One potential central engine capable of this is a highly magnetized, rapidly rotating neutron star, known as a magnetar. Magnetars can be formed by two compact objects coallescing, a scenario which is one of the leading progenitor models for short bursts in general. Assuming a magnetar is formed, we gain a value for the magnetic field and late time spin period for 9 of the extended emission bursts by fitting the magnetic dipole spin-down model of \citet{b39}. Assuming the magnetic field is constant, and the observed energy release during extended emission is entirely due to the spin-down of this magnetar, we then derive the spin period at birth for the sample. We find all birth spin periods are in good agreement with those predicted for a newly born magnetar.
\end{abstract}


\section{Introduction}

Gamma-Ray Bursts (GRBs) are the brightest phenomena in the Universe, releasing as much electromagnetic energy in tens of seconds as the entire Milky Way galaxy does in a few years \citep{b23}. They typically reach energies of around 5 x 10$^{50}$ ergs when beaming is accounted for \citep{b14}. Their temporal distribution shows a bimodality \citep{b18} which separates them into `long' or `short' GRBs (LGRB and SGRB respectively) depending on a parameter known as T$_{90}$; the time in which 90\% of the gamma-ray fluence is detected. Nominally, long bursts have T$_{90} > 2$ seconds, and short ones have T$_{90} < 2$ seconds, but in reality the distinction is far more blurred for a significant number of cases (eg \citealt{b17}; \citealt{b30}; \citealt{b30b}). Both classes have been observed to be distributed isotropically across the sky \citep{b5}. The most popular theory for SGRBs is that they are caused by  mergers of compact objects, such as double neutron star (NS--NS) binaries, NS -- black hole (BH) mergers, white dwarf (WD) -- NS mergers, WD--BH mergers or possibly even WD--WD mergers \citep{b28,b14a,b32,b3a,b8}. LGRBs are perhaps the better understood of the two classes, and are believed to be the product of massive star collapse \citep{b38,b29,b21} since, in cases where it would be observable, they are always accompanied by type Ib/c supernovae \citep{b15,b35}.

SGRBs are in fact not necessarily short. \citet{b26} found that 1/3 of their sample of SGRBs exhibit extended emission (EE) in their light curves, and an even greater fraction were found to be extended in the BATSE catalogue. EE is the term given to a rebrightening in the light curve after the initial emission spike. It happens at early times, usually beginning at $t \lesssim$ 10 s, and typically has lower peak flux but much longer duration than the initial spike, resulting in comparable total fluences between the two \citep{b30a}. Those bursts that were believed to exhibit EE were catalogued by \citet{b27}. These bursts present a challenge to the standard merger scenario, since they require an injection of energy arising seconds after the trigger, then naturally switching off at later times, around 100 s after trigger in the rest frame.

One central engine with the potential to provide such an energy supply is a rapidly spinning, highly magnetized neutron star, known as a magnetar \citep{b24,b6}. Whether by collapse or merger, this magnetar may be formed with sufficient rotational energy to avoid collapsing into a BH \citep{b10,b37,b39,b9}. GRBs require a rapidly rotating central engine with a strong, large scale magnetic field of around $10^{15}$ G or higher \citep{b22}, and a magnetar with such a field and an initial spin period $P_0 \approx 1$ ms has enough rotational energy to power a $\gtrsim 10^{52}$ erg GRB. Magnetar spin down has been often discussed as a source of GRBs, both long \citep{b39,b20,b8aa,b25,b3b} and short \citep{b13a,b33,b34}, in the literature, and has also been suggested as the origin of EE \citep{b24,b6} along with a variety of other mechanisms. Alternatives include a two-jet solution \citep{b2a}, fallback accretion \citep{b32a}, r-process heating of the accretion disc \citep{b24a} and magnetic reconnection and turbulence \citep{b40}. A major motivation for a common central engine can be seen in Figure 1; when all bursts are overlaid, a striking similarity can be seen in the evolution of the bursts, both temporally and energetically. Conformity like this is highly suggestive of a shared origin. \\

\begin{figure*}
\begin{center}
\includegraphics[width=8cm,angle=-90]{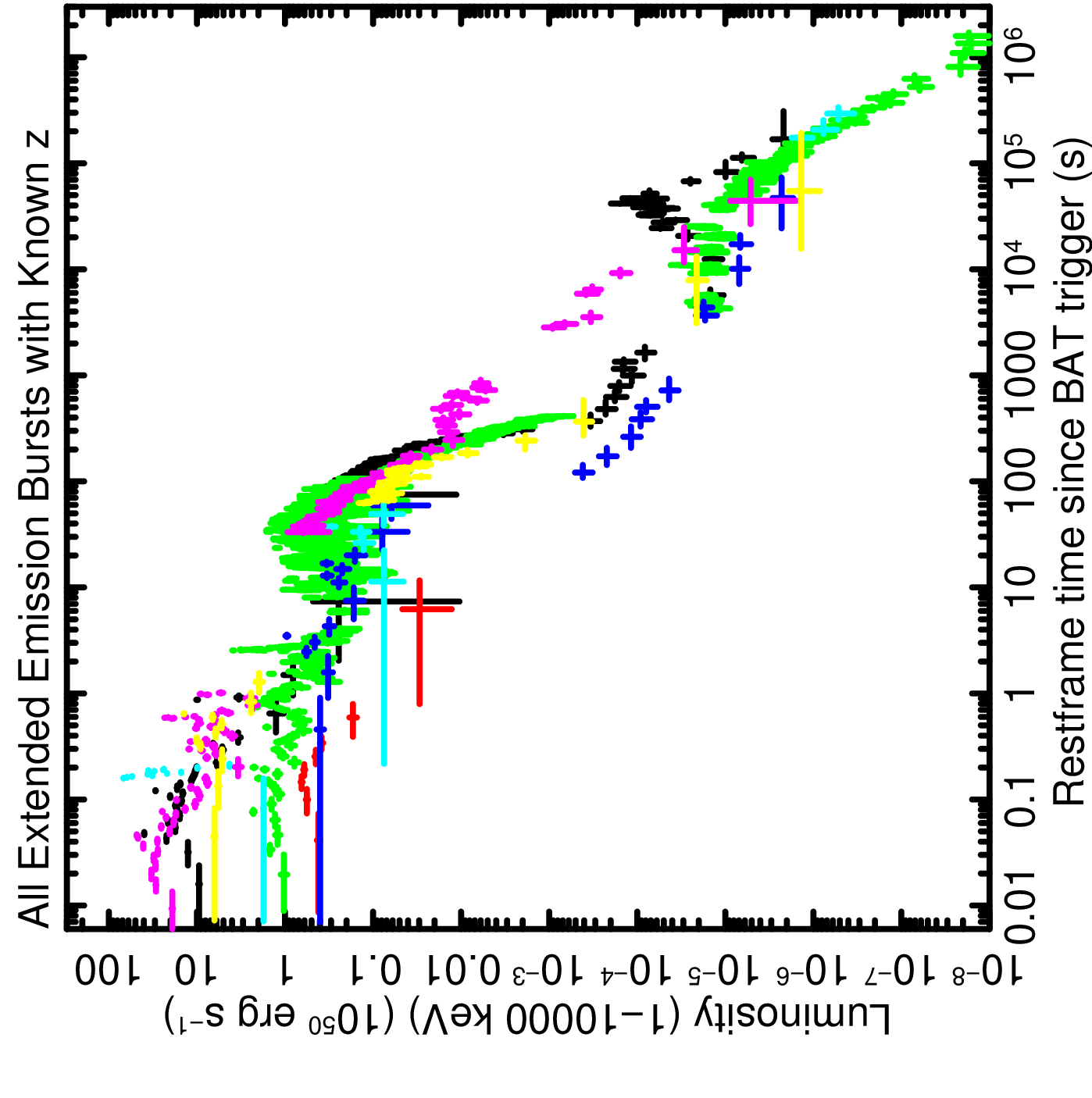}
\includegraphics[width=8cm,angle=-90]{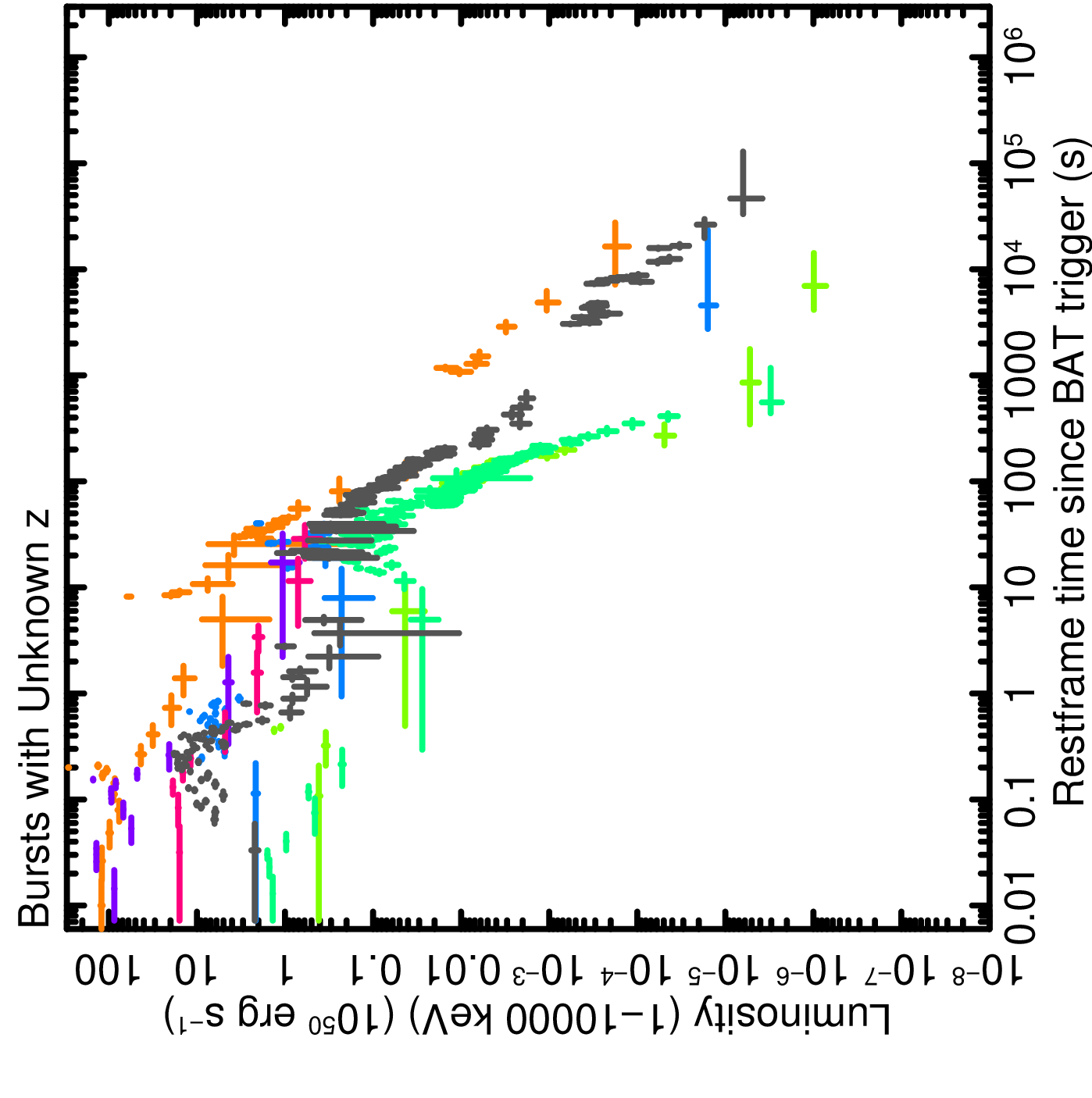}
\end{center}
\caption{Overlay of all bursts with extended emission, showing the apparently common evolutionary path. Left - Bursts with known $z$. Black - 050724; Red - 050911; Green - 060614; Blue - 061006; Light Blue - 061210; Pink - 070714B; Yellow - 071227. Right - Bursts using the sample average $z=0.39$. Orange - 051227 (Using the upper limit $z=2.8$, \citealt{b8b}); Lime Green - 080123; Mint Green - 080503; Blue - 090531B; Purple - 090715A; Red - 090916; Grey - 111121A.}
\end{figure*}

This paper is structured as follows:
 Section 2 contains the sample selection and data reduction process while section 3 details the motivation for finding a common central engine for EE GRBs. The magnetar model is introduced in section 4, the results are discussed in section 5, and the main conclusions are summarised in section 6.

\section{Sample Selection and Data Reduction}
\begin{table}
\begin{center}
\begin{tabular}{lccl}
\hline
GRB & $\Gamma$ & z & Ref.\\
\hline
050724 & 1.77 & 0.2576$^1$ & \citet{b7b} \\
050911 & 1.94 & 0.1646$^2$ & \citet{b29a} \\
051227 & 1.46 & 2.8$^{a,3}$ & \citet{b2aa} \\
060614 & 1.79 & 0.1254$^4$ & \citet{b30aa} \\
061006 & 2.03 & 0.4377$^5$ & \citet{b34a} \\
061210 & 2.20 & 0.4095$^6$ & \citet{b7aa} \\
070714B & 1.15 & 0.9224$^7$ & \citet{b30ca} \\
071227 & 1.54 & 0.381$^8$ & \citet{b34aa} \\
080123 & 1.99 & (0.39) & \citet{b36b} \\
080503 & 1.76 & (0.39) & \citet{b21a} \\
090531B & 2.07 & (0.39) & \citet{b8aaaa} \\
090715A & 1.38 & (0.39) & \citet{b30d} \\
090916 & 1.57 & (0.39) & \citet{b36a} \\
111121A & 1.50 & (0.39) & \citet{b8c} \\
\hline
\end{tabular}
\caption{Selected sample of EE GRBs. Bracketed values for redshift, $z$, indicate no published value was available. In these cases the mean value of the EE sample where $z$ is known was used. a - upper limit. 1 - \citet{b30c}; 2 - \citet{b3aa}; 3 - \citet{b8b}; 4 - \citet{b30ba}; 5 - \citet{b3ab}; 6 - \citet{b7a}; 7 - \citet{b17a}; 8 - \citet{b8a}}
\end{center}
\end{table}
The data used here were collected by the \emph{Swift} satellite \citep{b16}. Three instruments are carried on board: The Burst Alert Telescope (BAT; \citealt{b3}), which has an energy range of 15 -- 150 keV, the X-Ray Telescope (XRT; \citealt{b7}), energy range 0.3 -- 10 keV and the Ultra-Violet and Optical Telescope (UVOT; \citealt{b31}).

Raw BAT data for each burst were collected from the UK \emph{Swift} Science Data Centre (UKSSDC) archives and processed using the \emph{Swift} BAT pipeline tool {\sc batgrbproduct}. For all EE GRBs, we analysed the BAT data by creating lightcurves with a variety of binning in signal-to-noise ratios (SNR) and time, looking for evidence of EE at the 3$\sigma$ level where we consistently saw EE over more than 30 s. Using this method, a sample of 14 GRBs with EE was collected, including 12 which were identified as extended by \citet{b27}. This sample is shown in Table 1.

The XRT data were downloaded from the UKSSDC spectrum repository \citep{b12}, and were corrected for absorption using a ratio of (counts to flux unabsorbed)/(counts to flux observed). Details of the data reduction process can be found in \citet{b11,b12}. Standard {\sc heasoft} tools were used during data reduction.

To plot the BAT data alongside the XRT, the BAT light curves were extrapolated from their 15 -- 150 keV bandpass down to the XRT bandpass of 0.3 -- 10 keV using a correction factor comprised of the net count rate in the 15 -- 150 keV range and the extrapolated flux in the 0.3 -- 10 keV range, found using a power law fit to the pre-slew BAT spectrum in Xspec \citep{b1}. These combined light curves were made by taking the 4 ms BAT light curves from the {\sc batgrbproduct} pipeline and binning them with a SNR of 4, the one exception being GRB 080123, which was done with a SNR of 3. The light curves were then k-corrected, using the method described in \citet{b4} to give bolometric (1 -- 10000 keV) rest-frame light curves. The redshifts used during k-correction are displayed in Table 1. Where no constraints on redshift were available, the average for the sample, $z = 0.39$, was used. The value of $z = 2.8$ quoted for GRB 051227 is an upper limit \citep{b8b}.

\section{Evidence for a common central engine}
Figure 1 shows the EE sample from Table 1 plotted together. The left panel shows bursts with known redshift, whilst the right panel is the rest of the sample using the mean redshift value from bursts where $z$ is known. A striking similarity can be seen between the evolution of all EE bursts, particularly the ones where $z$ is known. The luminosity of the individual plateaus appear to be highly comparable between bursts, and the timescales in which these plateaus turn over also show a great deal of regularity. Such uniformity is highly suggestive of a common central engine, and hints at a unique difference between SGRBs and EE GRBs, but one that is common amongst the EE sample. One possible explanation for this uniformity is the correlation noted by \citet{b6} between magnetar outflow energy and jet opening angle, resulting in relatively constant isotropic power (within a factor $\sim 3$) for a given ejecta mass. GRB 051227 has been plotted in the right panel of Figure 1, since it does not have a firm redshift. Using $z = 2.8$ gives its EE tail (the 1st plateau at around $10 \lesssim t \lesssim 100$ seconds) a slightly higher luminosity than those in the left panel. \citet{b8b} give a tentative lower limit of $z \gtrsim 0.8$, and claim that the colour observations of the possible host galaxy are consistent with those of an irregular galaxy at $z \sim 0.8$. Using $z = 0.8$ would place GRB 051227 at around the same luminosity level as the known redshift bursts in Figure 1. We use $z = 2.8$ for this burst in the following analysis to place it at an extreme luminosity.

\begin{figure*}
\begin{center}
\includegraphics[width=5.8cm,angle=-90]{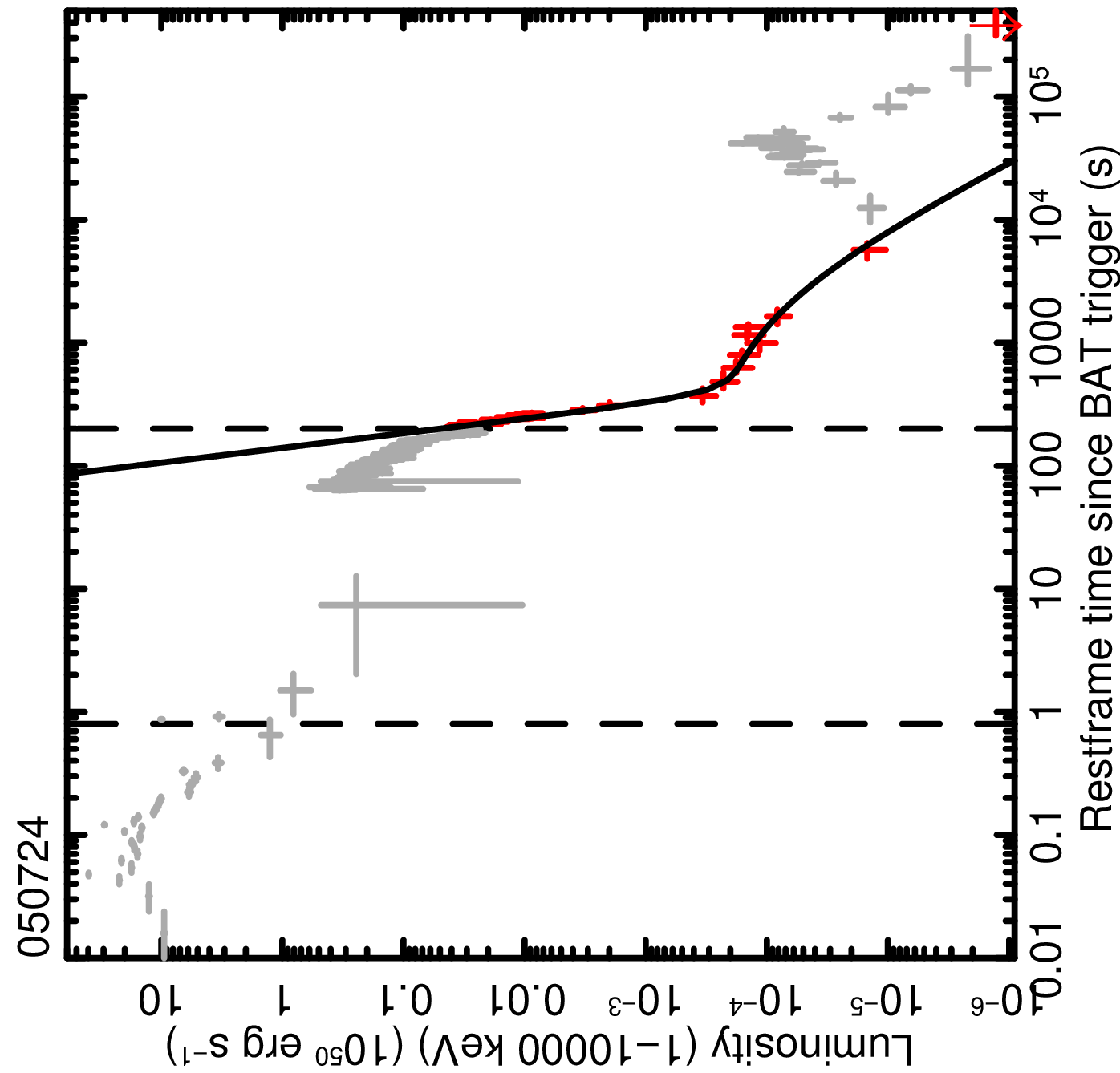}
\includegraphics[width=5.8cm,angle=-90]{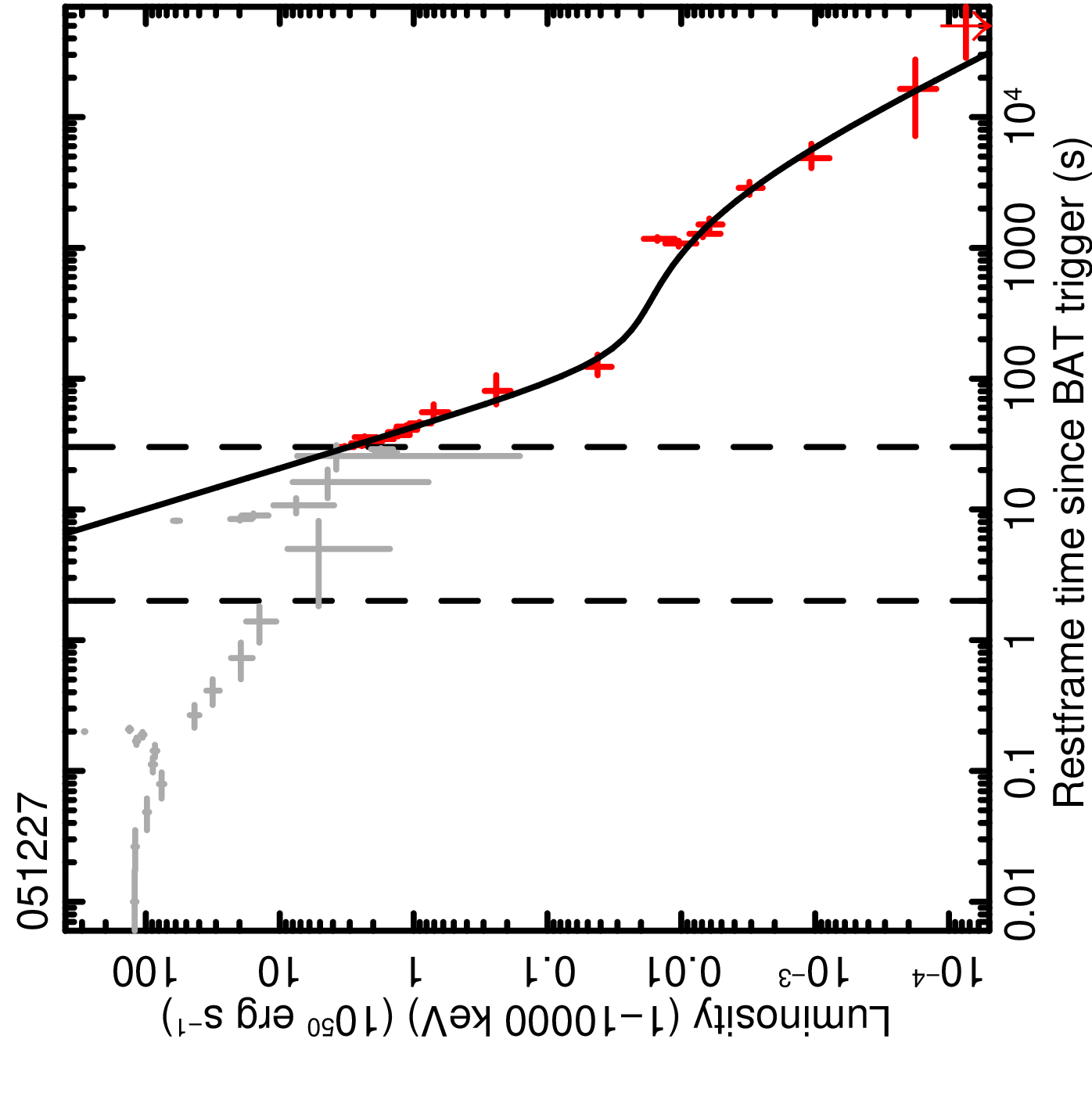}
\includegraphics[width=5.8cm,angle=-90]{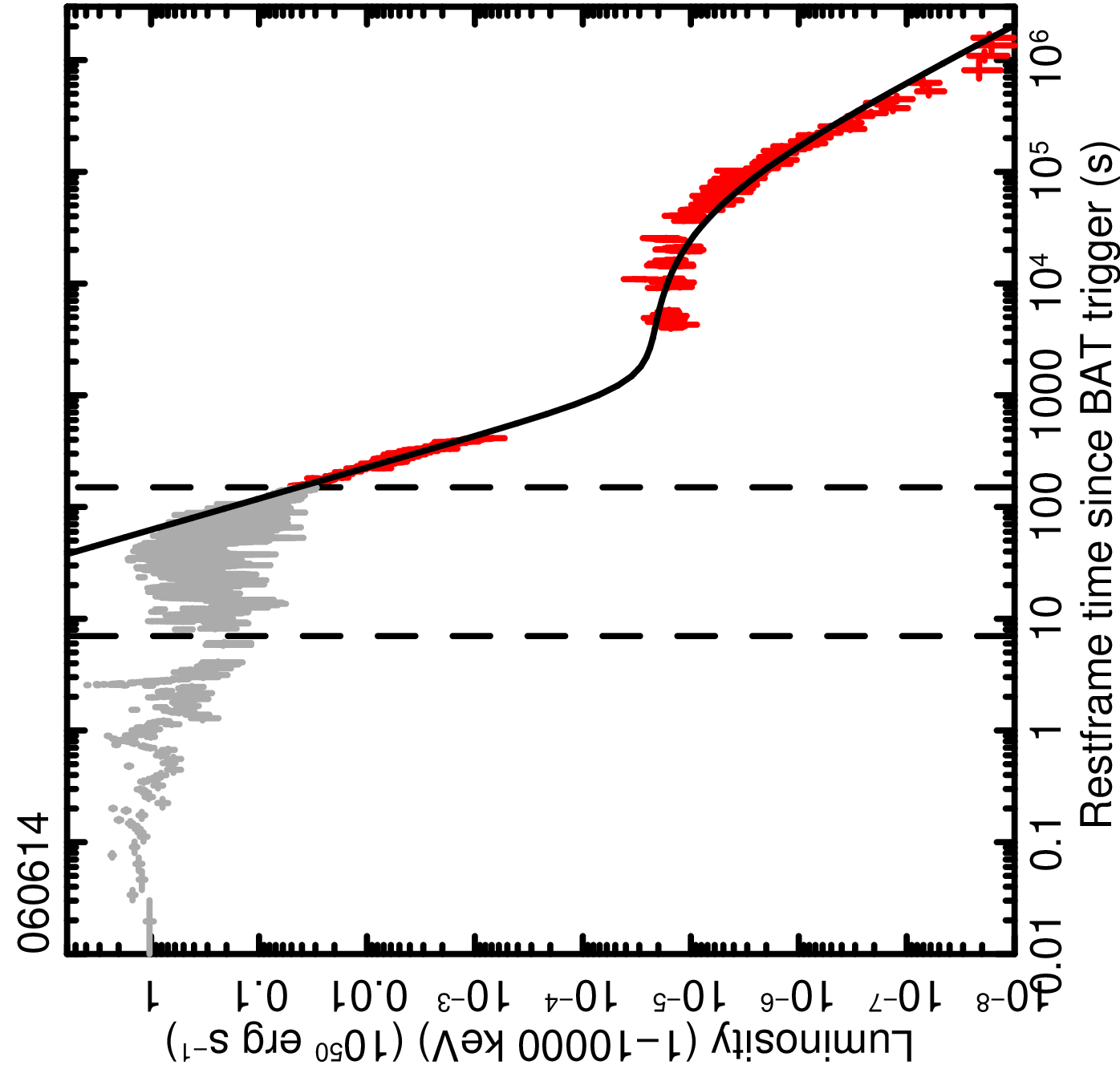}
\includegraphics[width=5.8cm,angle=-90]{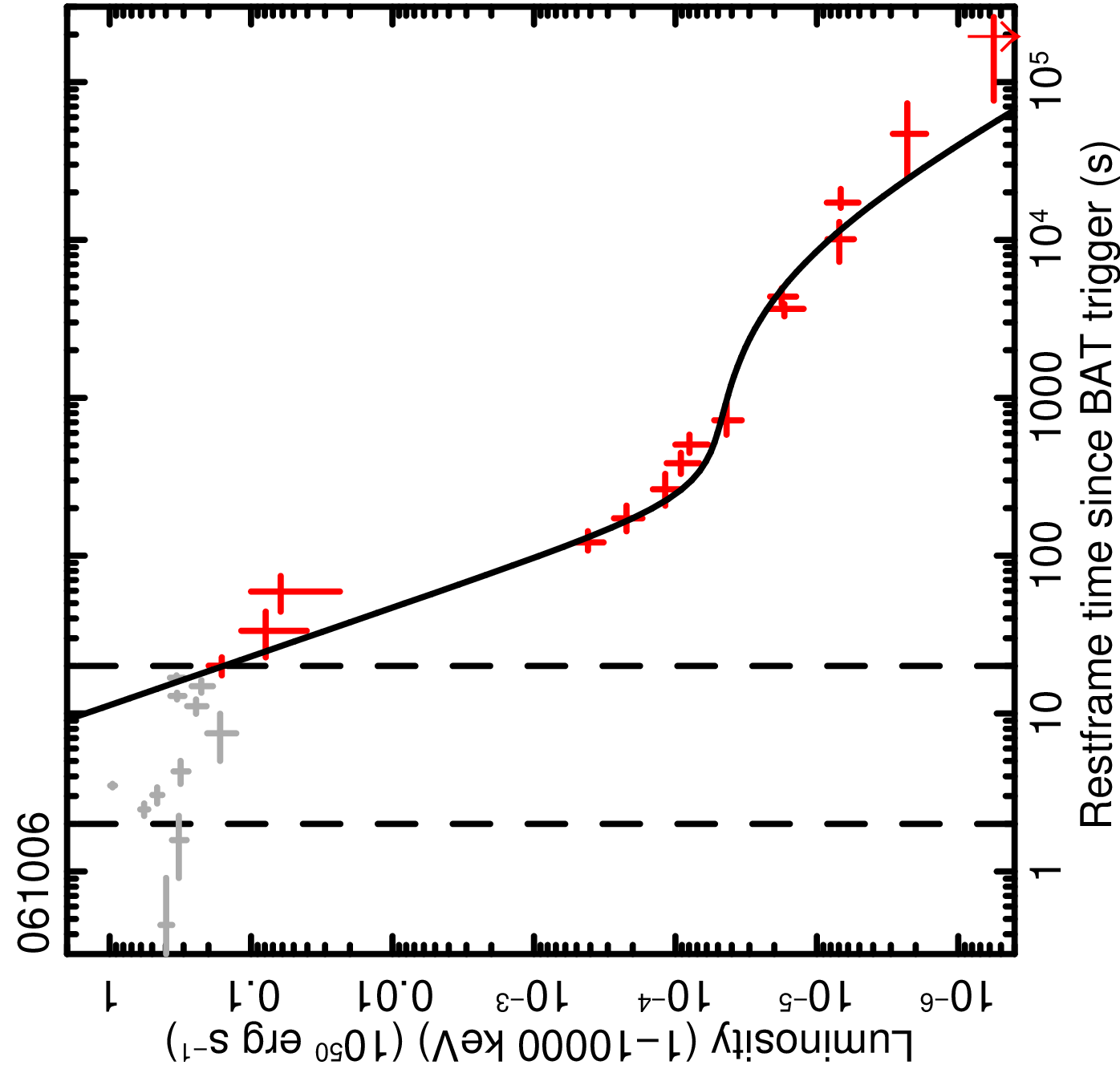}
\includegraphics[width=5.8cm,angle=-90]{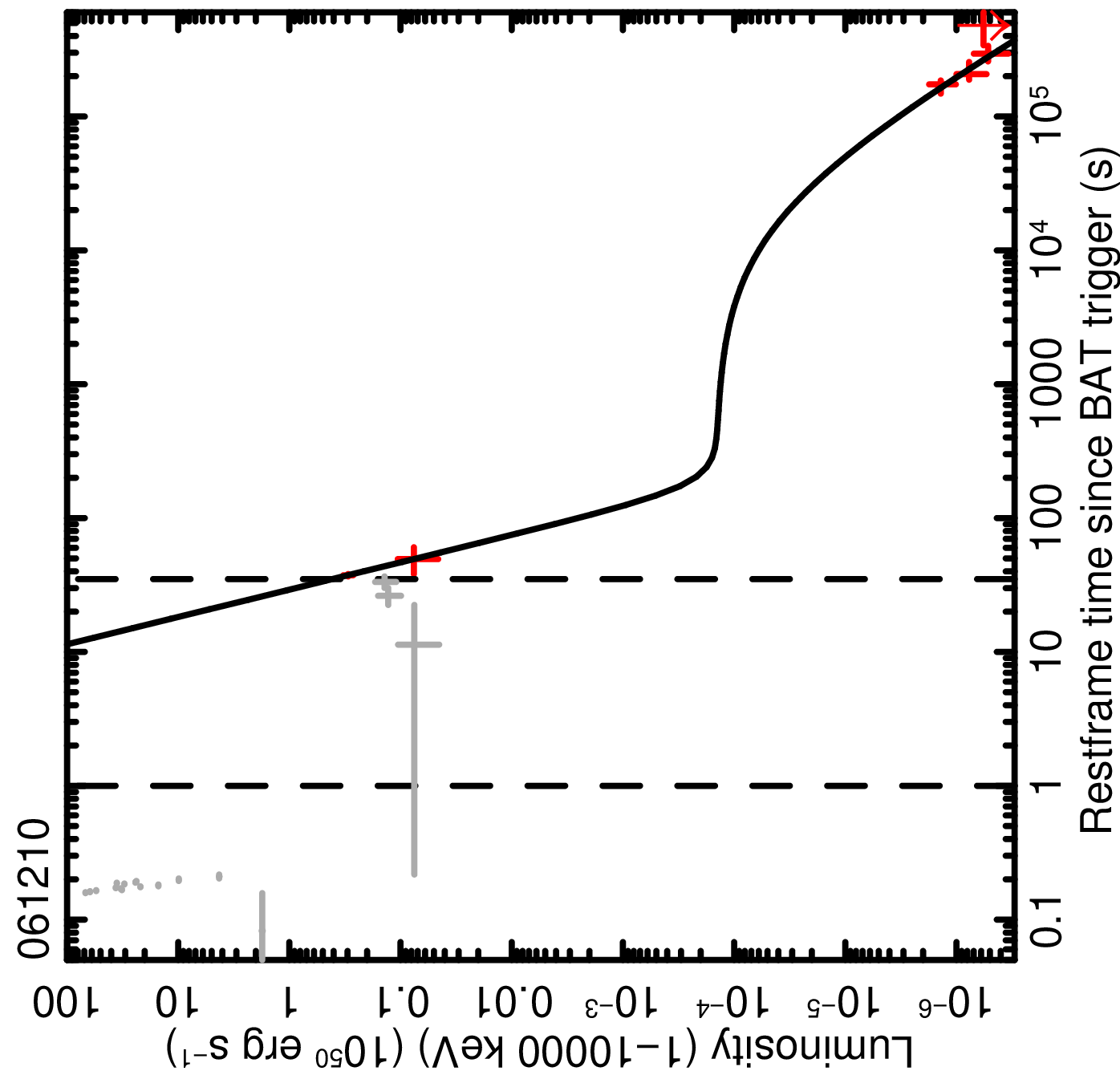}
\includegraphics[width=5.8cm,angle=-90]{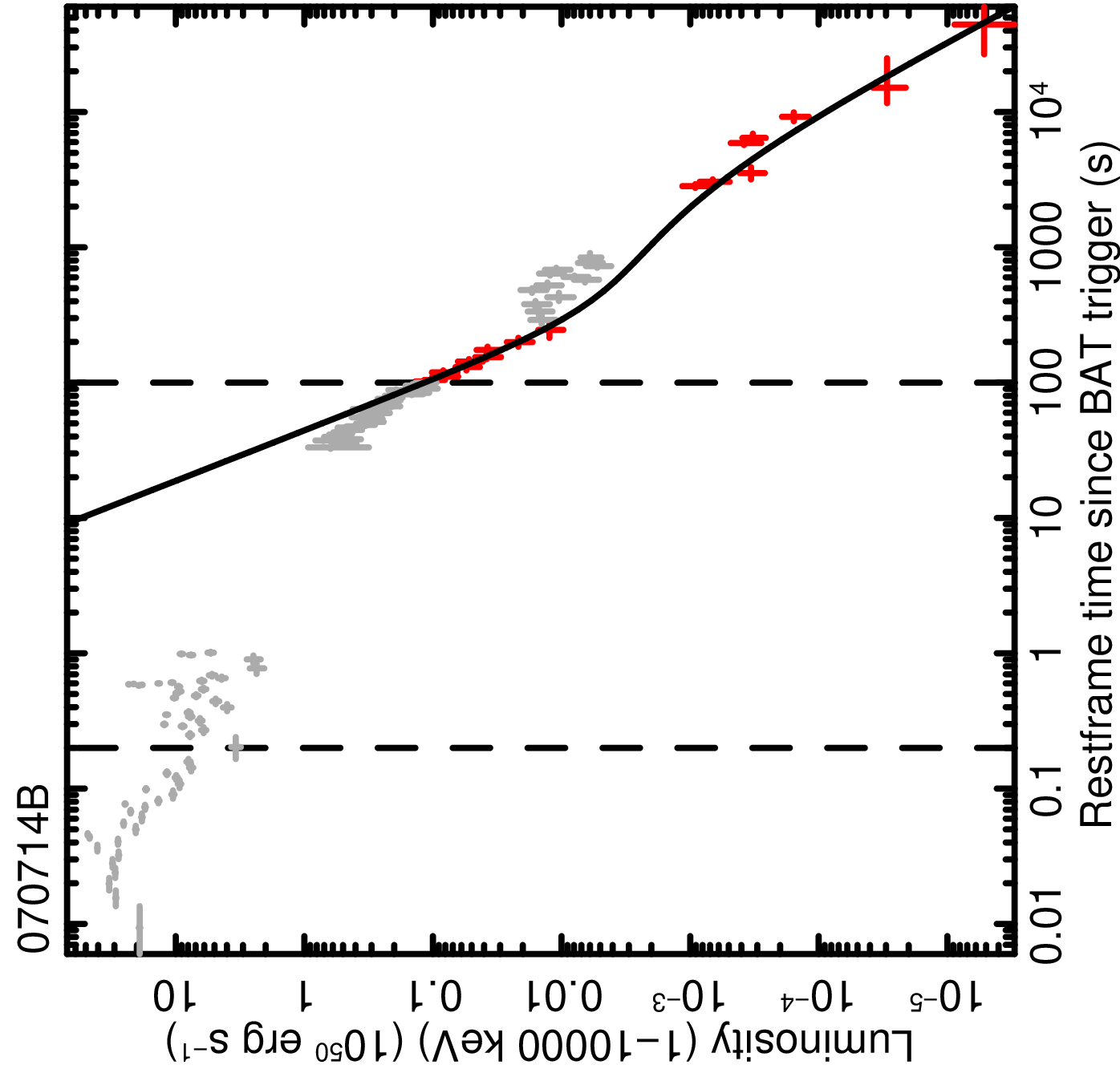}
\includegraphics[width=5.8cm,angle=-90]{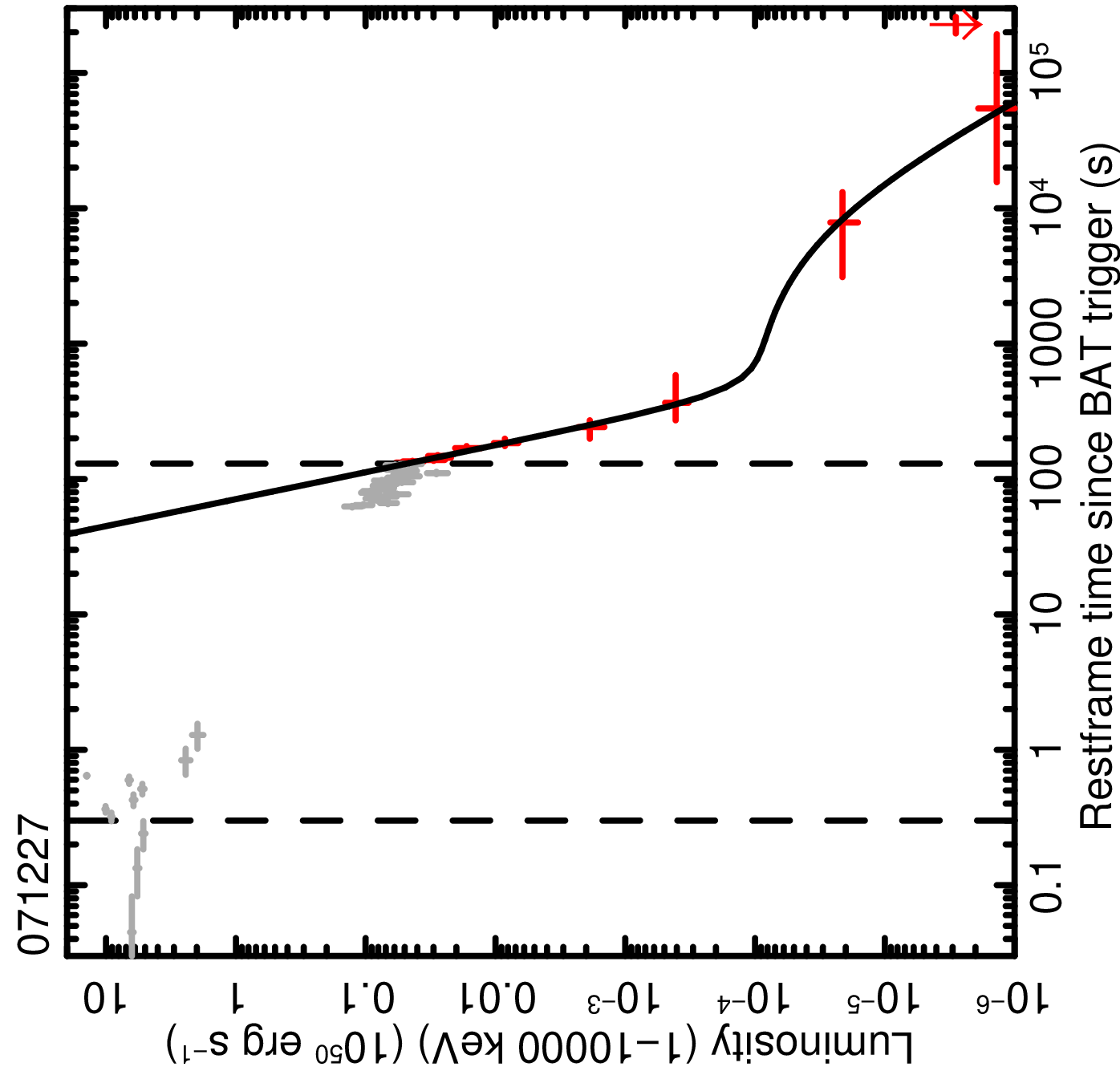}
\includegraphics[width=5.8cm,angle=-90]{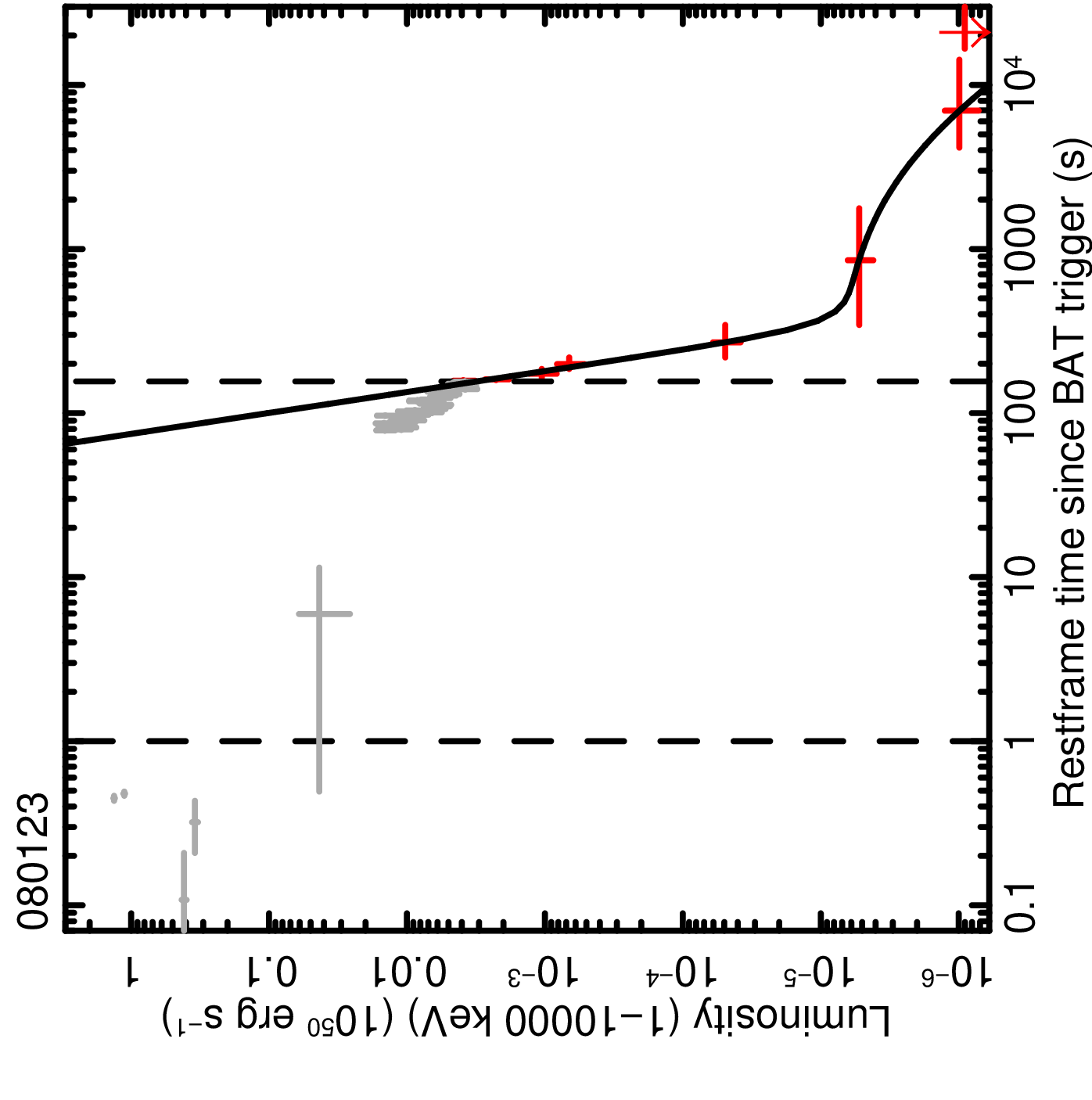}
\includegraphics[width=5.8cm,angle=-90]{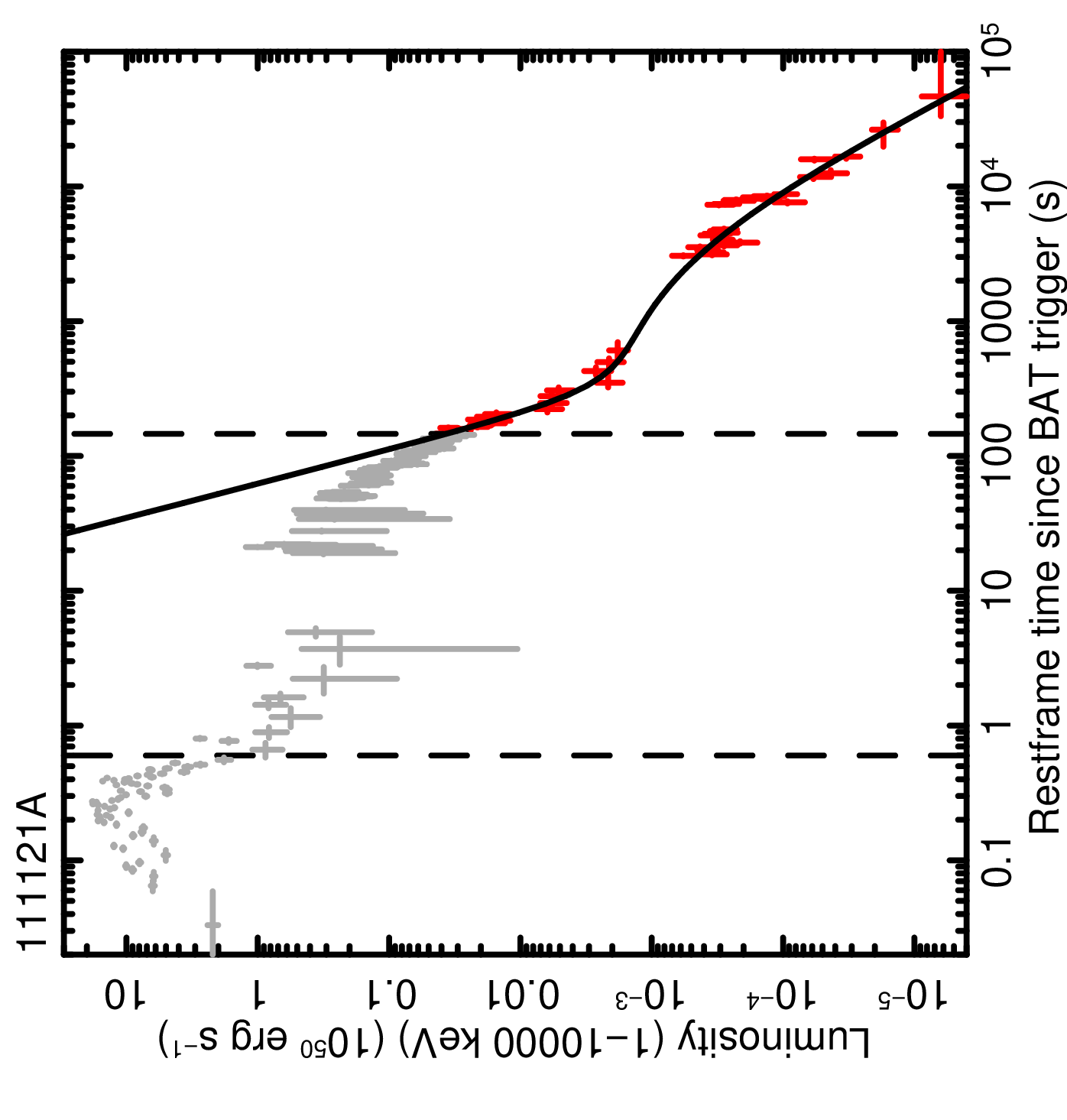}
\end{center}
\caption{Light curves fit with the magnetic dipole spin-down model. Red points have been fitted to, grey points have not, most noticably the late-time flare in GRB 050724 and the $\sim 400$ s flare in GRB 070714B. The vertical dashed lines indicate the extended emission region, between which extended emission energy is calculated by integrating under the curve.}
\end{figure*}

\section{The Magnetar Model}
\subsection{Magnetic dipole spin-down}
The magnetic dipole spin-down model is detailed in \citet{b39}, and has been used on both SGRBs (eg \citealt{b13a}; \citealt{b34}) and LGRBs (eg \citealt{b20}; \citealt{b8aa}; \citealt{b3b}). The model is fitted to the the late-time plateau, seen emerging from beneath the fading EE tail in Figure 2 at times of around $100$ -- $1000$ s. This allows the magnetic field and spin period of the central magnetar to be derived, although the calculated spin period must then be corrected for spin-down during EE to get the true birth period (see section 4.2).

The basic outline is that the central engine, in this case a magnetar, emits both an initial impulse energy $E_{imp}$ as well as a continuous injection luminosity which varies as a power-law in the emission time. The initial impulse energy represents the prompt emission of the burst (excluding EE), and is a short, violent event which transitions into a power-law decay at very early times. The continuous injection luminosity is the product of the magnetar spinning down, and begins as soon as the magnetar is formed. Although it is present throughout, it's at a much lower level than the initial impulse, and so is initially hidden beneath the more luminous component. At a critical time, $T_c$, the prompt emission has faded enough so that the injection luminosity begins to dominate the light curve, causing it to flatten. This effect can be seen in the red datapoints in Figure 2. The plateau then re-steepens after the characteristic timescale for dipole spin-down, $T_{em}$. At this point, the magnetar reveals itself as either unstable, collapsing into a BH with a sudden drop in the light curve, or stable, continuing to decay with a comparatively shallow power-law. For this plateau to appear, $T_{em}$ must be greater than $T_c$, otherwise the continuous injection luminosity is spent before the prompt emission has faded sufficiently for it to be observable.

To derive the parameters that control the injection luminosity plateau, the dimensions of the plateau itself must be ascertained by fitting. The area of interest for fitting is the point at which the continuous injection (dipole spin-down) luminosity emerges from beneath the initial impulse energy and the fading EE tail, shown by the red datapoints in Figure 2. Obtaining fits that describe the luminosity and duration of this plateau allows the magnetic field and spin period of the sample to be found. The key equations for the model are:
\begin{eqnarray}
T_{em,3} = 2.05\, (I_{45}B_{p,15}^{-2}P_{0,-3}^2R_6^{-6})
\\
L_{0,49} \sim (B_{p,15}^2P_{0,-3}^{-4}R_6^6)
\\
B_{p,15}^2 = 4.2025I_{45}^2R_6^{-6}L_{0,49}^{-1}T_{em,3}^{-2}
\\
P_{0,-3}^2 = 2.05I_{45}L_{0,49}^{-1}T_{em,3}^{-1}
\end{eqnarray}
where $T_{em,3}$ is the characteristic timescale for dipole spin-down in 10$^3$ s, $L_{0,49}$ is the plateau luminosity in 10$^{49}$ erg s$^{-1}$, $I_{45}$ is the moment of inertia in units of 10$^{45}$ g cm$^2$, $B_{p,15}$ is the magnetic field strength at the poles in units of 10$^{15}$ G, $R_6$ is the radius of the neutron star in 10$^6$ cm and $P_{0,-3}$ is the spin period of the magnetar in milliseconds. The mass of the magnetar was set to $1.4$ $M_{\odot}$ and the radius was $10^6$ cm. Using these values, the moment of inertia, I, is $9.75 \times 10^{44}$ g cm$^2$. Equations 1 -- 4 are taken from \citet{b39} and were combined into a {\sc qdp} COmponent Definition (COD) file for fitting to data by \citet{b34} during their work. This COD file was used to obtain fits as previously in the current work. It has been assumed that emission is both isotropic and 100\% efficient, since little is known about the precise emission mechanism and beaming angle. \citet{b20} discussed the effects of beaming in the context of the magnetar model, and showed that a narrower opening angle results in higher $B$ and $P$ (slower spin). This is illustrated by their Figure 4.

The magnetic dipole spin-down model was fitted to the late time data of the rest-frame light curves of 9 GRBs with EE. Of the original sample of 14 bursts, 5 did not contain sufficient datapoints for accurate model fitting and were dropped from the sample. GRB 050911, GRB 090715A and GRB 090916 do not have XRT data available, and the XRT data for GRB 090531B contains only a single point and an upper limit. GRB 080503 either has an incredibly weak dipole plateau or none at all \citep{b30a}, so values for magnetic field and spin period were unobtainable. Table 2 contains the results of the fitting to the 9 remaining GRBs.

\begin{table}
\begin{center}
\begin{tabular}{lccccc}
\hline
GRB & Region & P$_0$ & B & $\alpha$ & Reduced \\
 & (s) & (ms) & ($10^{15}$G) & & $\chi^2$ \\
\hline
050724 & $\geq 200$ & 19.5$^{+1.10}_{-0.97}$ & 21.2$^{+3.77}_{-3.01}$ & 8.39$^{+0.01}_{-0.01}$ & 2.68 \\
051227 & $\geq 30$ & 2.34$^{+0.14}_{-0.12}$ & 2.82$^{+0.33}_{-0.29}$ & 3.20$^{+0.22}_{-0.18}$ & 1.04 \\
060614 & $\geq 150$ & 14.0$^{+0.14}_{-0.13}$ & 3.10$^{+0.06}_{-0.05}$ & 3.59$^{+0.04}_{-0.04}$ & 1.44 \\
061006 & $\geq 20$ & 24.2$^{+1.39}_{-1.25}$ & 14.1$^{+2.60}_{-2.44}$ & 3.24$^{+0.17}_{-0.19}$ & 2.12 \\
061210 & $\geq 35$ & 8.89$^{+4.55}_{-5.78}$ & 3.04$^{+0.36}_{-0.28}$ & 4.90$^{+0.03}_{-0.03}$ & 0.57 \\
070714B & $\geq 100$ & 5.14$^{+0.68}_{-0.75}$ & 6.04$^{+0.68}_{-0.61}$ & 2.69$^{+0.43}_{-0.31}$ & 1.31 \\
071227 & $\geq 130$ & 16.9$^{+2.38}_{-2.43}$ & 9.62$^{+3.56}_{-2.33}$ & 5.02$^{+0.54}_{-0.30}$ & 0.57 \\
080123 & $\geq 156$ & 82.5$^{+9.44}_{-7.35}$ & 60.6$^{+19.0}_{-13.4}$ & 7.85$^{+0.02}_{-0.02}$ & 1.94 \\
111121A & $\geq 146$ & 6.15$^{+0.16}_{-0.19}$ & 5.70$^{+0.24}_{-0.27}$ & 3.94$^{+0.41}_{-0.36}$ & 1.27 \\
\hline
\end{tabular}
\caption{Results of fitting the magnetic dipole spin-down model to the sample of extended emission bursts. $P_0$ is the spin period after EE in ms, $B$ is the magnetic field in $10^{15}$G. $\alpha$ is the power-law of the decay slope. All errors are 1$\sigma$.}
\end{center}
\end{table}

Figure 2 shows the individual fits for each of the 9 bursts, along with the estimated EE region, denoted by the vertical dashed lines. The start of the EE region is taken as the first upturn in the light curve after the initial prompt emission spike. EE is said to have ceased at the time of the final power-law decay before the onset of the magnetic dipole spin-down plateau. Using these definitions, we were able to reasonably recreate the fluence ratios of \citet{b30a} and the EE duration times of \citet{b27}. For each burst, a solution was found in which the data was accurately traced by the model, and the results returned for the values of $B$ and $P_0$ lie unambiguously in allowed parameter space.

$P_0$ is refered to as the initial spin period of the magnetar by \citet{b34}. Whilst this is true for short bursts where spin down only occurs due to EM dipole radiation, the story is more complicated for EE bursts. Since we are assuming the extraction of rotational energy from the spin of the magnetar is the mechanism behind the EE tail, the spin period during this time must be variable. In fact, during this time the magnetar may be spun up by accretion on to the surface, or down by a variety of mechanisms in addition to the constant dipole spin down that exists in the pure short GRB case. Thus, for these EE bursts, $P_0$ has been taken as the spin period after EE. We return to this issue in section 4.2.

The derived values of $B$ and $P_0$ are plotted against each other in Figure 3, where the 3 vertical and 2 horizontal lines denote allowed parameter space for the birth of a magnetar powering a GRB. Our lower limit on spin period is the spin break-up frequency for a $1.4$ M$_{\odot}$ NS with a radius of $10$ km \citep{b19}. Also plotted is the limit for a $2.1$ M$_{\odot}$ NS with the same radius, shown by the dashed line. These limits may vary with uncertainties in the equation of state of the NS. \citet{b37} calculated the minimum allowed spin frequency at birth if the progenitor is the accretion induced collapse of a WD. Based on conservation of angular momentum, the upper spin period limit would be $10$ ms for this type of progenitor. The minimum magnetic field required to produce a GRB observable in the gamma band \citep{b36}, sets the lower boundary for $B$ at $10^{15}$ G. The initial impulse energy of the burst is accounted for by a power-law with a decay slope $\alpha$ after the prompt emission. In practice, this power-law simply models the light curves in the region between the EE tail and the dipole spin-down plateau. It can be seen from the results and the fits in Figure 2 that all magnetars in this sample are stable.

\begin{figure}
\begin{center}
\includegraphics[width=8cm,angle=-90]{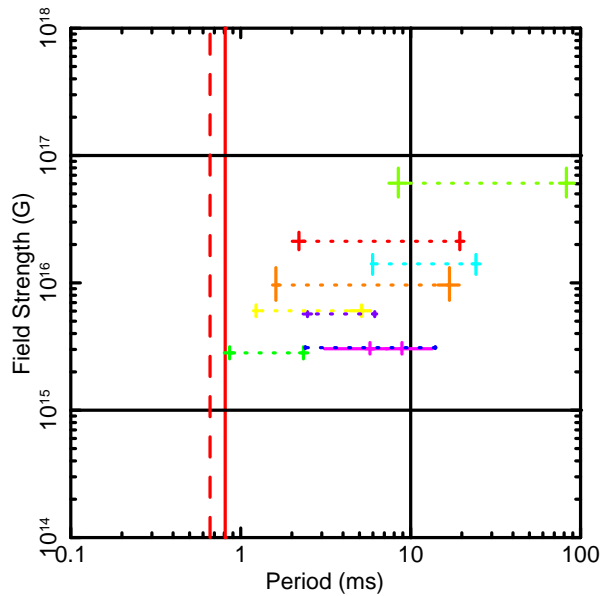}
\end{center}
\caption{Plot of the spin period before and after extended emission against magnetic field strength. Spin period evolves from the left ($P_i$, birth) of the dotted lines, through extended emission, to the right ($P_0$). Limits (denoted by solid lines): Vertical left (red), spin break-up frequency for a $1.4$ M$_{\odot}$ (solid) and $2.1$ M$_{\odot}$ (dashed) NS with a $10$ km radius \citep{b19}; Vertical right (black), minimum allowed spin frequency at birth, based on conservation of angular momentum during the accretion induced collapse of a WD \citep{b37}; Horizontal lower, minimum magnetic field required to produce a GRB observable in gamma band \citep{b36}. Bursts: Red - 050724; Green - 051227; Dark Blue - 060614; Light Blue - 061006; Pink - 061210; Yellow - 070714B; Orange - 071227; Light Green - 080123; Purple - 111121A.}
\end{figure}

\subsection{The extended emission tail}
Once a fit has been found for the late time data of a specific burst, the magnetic field strength, $B$, and the spin period after EE, $P_0$ become known quantities. The energy release of the EE tail can be calculated fairly simply by estimating the points on the light curve where EE begins and ends and integrating under the curve between these two points, ie $\mathrm{d}E = L\,\mathrm{d}t$. This is done using linear interpolation between points, and the calculated EE energies are displayed in Table 3. Assuming a constant magnetic field, and that energy injection during the EE period is entirely from the spin-down emission of the magnetar, the spin period the magnetar possessed at birth, $P_i$, can be calculated using
\begin{equation}
\Delta E = 2\pi^2I(P_i^{-2} - P_0^{-2})
\end{equation}
where $\Delta E$ is the energy in the EE tail, $I$ is the moment of inertia, $P_0$ is the spin period of the magnetar after EE and $P_i$ is the birth spin period. Table 3 contains the results from this process, including the time boundaries for EE, the energy found by integration, and the resultant value derived for $P_i$.

\begin{table}
\begin{center}
\begin{tabular}{lcccc}
\hline
GRB & $T_{start}$ (s)& $T_{stop}$ (s)& $\Delta E$ (10$^{50}$erg) & $P_i$ (ms) \\
\hline
050724 & 0.8 & 200 & 39.4$\pm 6.74$ & 2.20$\pm 0.19$ \\
051227 & 2 & 30 & 224$\pm 30.9$ & 0.86$\pm 0.06$ \\
060614 & 7 & 150 & 32.2$\pm 0.35$ & 2.40$\pm 0.01$ \\
061006 & 2 & 20 & 5.05$\pm 0.16$ & 5.97$\pm 0.10$ \\
061210 & 1 & 35 & 3.37$\pm 0.42$ & 5.76$\pm 0.36$ \\
070714B & 0.2 & 100 & 121$\pm 7.02$ & 1.23$\pm 0.04$ \\
071227 & 0.3 & 130 & 73.8$\pm 6.08$ & 1.61$\pm 0.07$ \\
080123 & 1 & 156 & 2.66$\pm 0.64$ & 8.46$\pm 1.01$ \\
111121A & 0.6 & 146 & 26.3$\pm 2.90$ & 2.47$\pm 0.14$ \\
\hline
\end{tabular}
\caption{Results for the birth spin period, $P_i$, derived from the extended emission energy, $\Delta E$. $T_{start}$ and $T_{stop}$ mark the beginning and end of the extended tail where the energy is estimated. All errors are 1$\sigma$.}
\end{center}
\end{table}

\section{Discussion}
The calculated spin periods for the birth of the magnetar lie comfortably within allowed parameter space (Figure 3) and are consistent with values predicted in the literature \citep{b37,b35a,b8}. Bursts that do not have a set redshift may vary on the energy scale, with an error of $0.5$ in $z$ roughly corresponding to an order of magnitude in the luminosity scale. \citet{b33,b34} discussed the effect of varying redshift on the results for $B$ and $P_0$ in their work, and the argument is well illustrated by Figure 9(b) in \citet{b34}. The general result is that a higher $z$ corresponds to a lower rotation period (ie faster spin) and lower magnetic field. A good example is the change in results if the sample average redshift $z = 0.39$ is used for GRB 051227; fitting the magnetic dipole spin-down model then gives a magnetic field $B = 22.0^{+2.54}_{-2.27} \times 10^{15}$ G and a spin period of $P_0 = 30.2^{+1.79}_{-1.59}$ ms. The light curve is also far less luminous. The EE energy release is just $\Delta E = 1.34 \pm 0.19 \times 10^{50}$ ergs, which translates into $P_i = 11.1 \pm 0.77$ ms.

Figure 4 shows where the values found for $B$ and $P_i$ place the EE bursts relative to other SGRB and LGRB populations taken from Figure 9(a) of \citet{b34}. It can be seen that the EE bursts show properties that most closely resemble the unstable magnetar population of SGRBs. Since both magnetic field and spin period are very similar between these two groups, the difference must lie in some other property, perhaps mass or formation mechanism. This key difference must prevent the EE sample bursts from collapsing into BHs, and enable, perhaps even cause, the release of EE energy. \citet{b32a} showed that accretion discs and fallback accretion exhibit a much wider spread of behaviours when the compact objects involved in the merger have different masses. In their work, a NS -- NS binary showed fairly homogeneous behaviour, whilst a NS -- BH merger produced a much broader spread of fallback activity. A magnetar cannot be formed from a BH, but the same principle of unequal masses can be achieved by a system involving a NS -- WD merger, or, with the discovery of increasingly \citep{b8d}, possibly a more exotic NS -- NS system.

\begin{figure}
\begin{center}
\includegraphics[width=8cm,angle=-90]{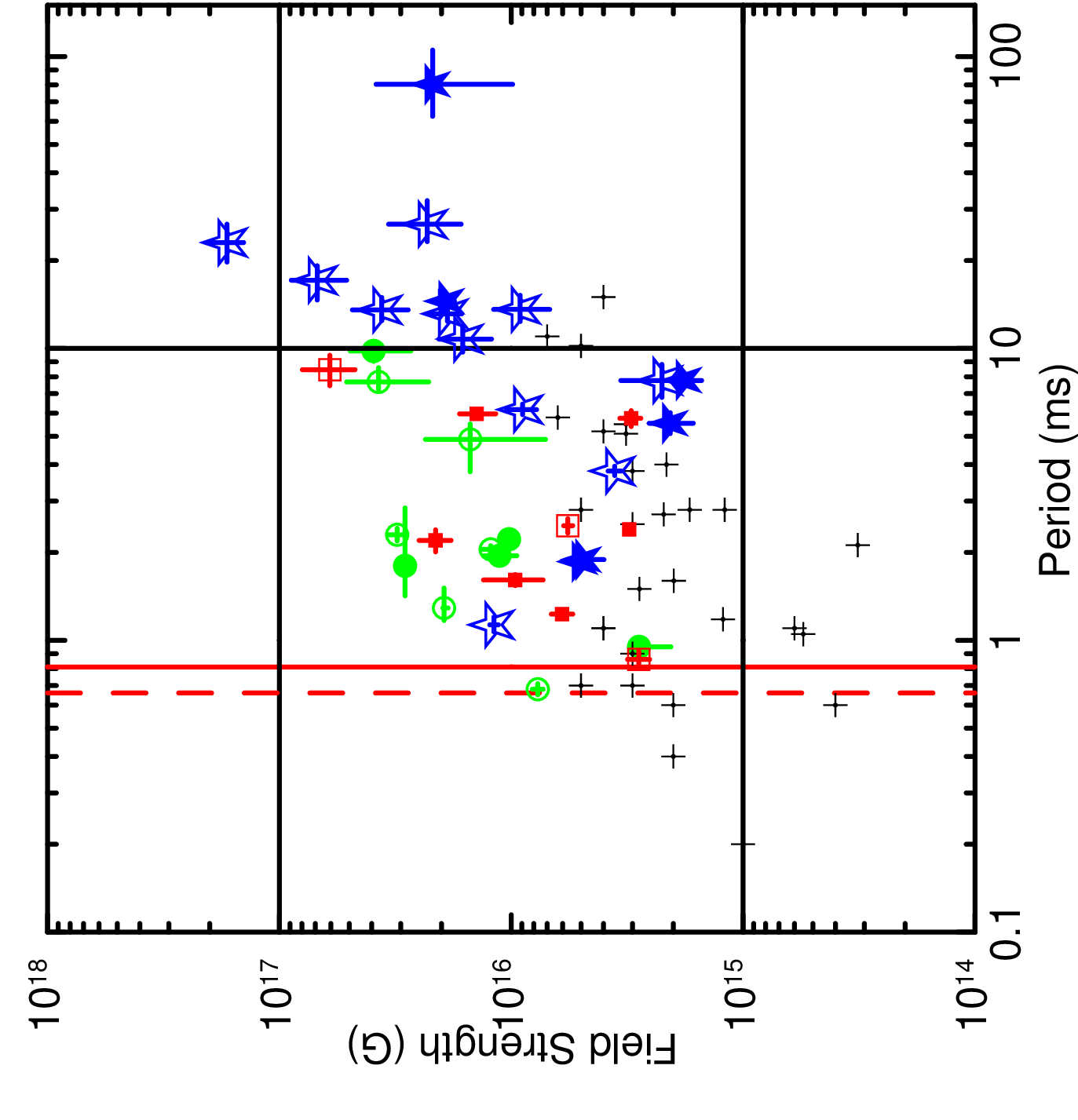}
\end{center}
\caption{A plot of magnetic field strength versus spin period. The solid (dashed) red line represents the spin break up period for a collapsar (binary merger) progenitor \citep{b19}. Blue stars: stable magnetars and green circles: unstable magnetars which collapse to form a BH \citep{b34}. Black `+' symbols are the LGRB candidates identified by \citet{b20,b8aa,b3b}. The red squares are the magnetic fields and birth spin periods ($P_i$) of the present work. Filled symbols have observed redshifts, open symbols use the sample average redshift, which is $z = 0.39$ for extended bursts and $z = 0.72$ for the short bursts from \citet{b34}.}
\end{figure}

\section{Conclusions}
EE GRB light curves show a remarkable uniformity when plotted alongside each other, particularly amongst the bursts where redshift is known. This consistency in plateau luminosity and turnover times suggests EE GRBs share a common progenitor mechanism which distinguishes them from ordinary SGRBs.

We have fitted the magnetic dipole spin-down model of \citet{b39} to the late-time data of the light curves of 9 GRBs under the assumption that the central engine is a highly magnetized neutron star. These fits have yielded values for the magnetic field strength and late-time spin period. We have also performed calculations of the energy contained in the EE region of bursts in this sample. Assuming this energy release is due to the spin-down of the central magnetar, and assuming a constant magnetic field, we infer the spin period these magnetars possessed at birth. The spin periods found are in good agreement with published values for the birth of a magnetar (eg \citealt{b35a}; \citealt{b8}; \citealt{b37}). These results are consistent with the idea that EE GRBs could be powered by a spinning-down magnetar.

\section{Acknowledgements}
BG acknowledges funding from the Science and Technology Funding Council. The work makes use of data supplied by the UK \emph{Swift} Science Data Centre at the University of Leicester and the \emph{Swift} satellite. \emph{Swift}, launched in November 2004, is a NASA mission in partnership with the Italian Space Agency and the UK Space Agency. \emph{Swift} is managed by NASA Goddard. Penn State University controls science and flight operations from the Mission Operations Center in University Park, Pennsylvania. Los Alamos National Laboratory provides gamma-ray imaging analysis. We thank the anonymous referee for their swift response.

\label{lastpage}

\end{document}